\documentclass[conference]{IEEEtran}
\usepackage[]{graphicx}
\usepackage{epstopdf}
\usepackage{ctable,amsmath}
\newcommand{\comments}[1]{}

\begin{document}

\title{Adaptive Frequency Cepstral Coefficients for Word Mispronunciation Detection}
\author{\IEEEauthorblockN{Zhenhao Ge, Sudhendu R. Sharma, Mark J.T. Smith}
\IEEEauthorblockA{School of Electrical and Computer Engineering \\
  Purdue University, West Lafayette, Indiana, 47907, USA}}
  
\maketitle

\begin{abstract}
Systems based on automatic speech recognition (ASR) technology can provide important functionality  in computer assisted language learning  applications. This is a young but growing area of research motivated by the large number of students studying foreign languages. Here we propose a Hidden Markov Model (HMM)-based method to detect mispronunciations. Exploiting the specific dialog scripting employed in language learning software, HMMs are trained for different pronunciations. New adaptive features have been developed and obtained through an adaptive warping of the frequency scale prior to computing the cepstral coefficients. The optimization criterion used for the warping function is to maximize separation of two major groups of pronunciations (native and non-native) in terms of classification rate. Experimental  results show that the adaptive frequency scale yields a better coefficient representation leading to higher classification rates in comparison with conventional HMMs using Mel-frequency cepstral coefficients.
\end{abstract}


\section{Introduction}

In this paper, we consider the problem of pronunciation detection and evaluation. This is an important challenge for which DSP algorithms can provide cost-effective solutions. Millions of people study foreign languages. Some are fortunate to have one-on-one time with the instructor, where they receive correction on words they mispronounce.  Most students, however, do not. Since the number of language students out number instructors by factors of 20, 30, or more, instructors cannot effectively spend significant amounts of time with students individually. 

Software tools, like Rosetta Stone and TellMeMore, can help address instructor access limitations by providing learners with visual aids such as waveform displays, plots of pitch contours, and spectrograms.  However, all fall short of having an instructor point out mispronunciations and having the student repeat the utterances until he or she pronounces them   correctly.   The challenge for the DSP algorithm is to recognize mispronunciations and provide feedback to the learners, just as a teacher would do. The work discussed in this paper addresses this problem for isolated words. Though the method is language independent, we are currently focusing on Spanish---specifically American students learning Spanish.

The problem of mispronunciation detection has gained significant interest in the the last two decades. Mispronunciation detection shares much in common with automatic speech recognition (ASR) and majority of mispronunciation detection systems use statistical models such as Hidden Markov Models (HMMs) in order to detect mispronunciations. Works by \cite{Bernstein,Ronen,Neumeyer,Franco,Zhang,Chen} have used HMMs to score pronunication quality based on different measures in order to detect mispronunciations. More recent works by \cite{Wang,Harrison1,Qian,Lo} have attempted to improve the detection accuracy by using extended recongnition networks that incorporate cross-language phonological rules, i.e. rules that dictate how a learner's first language affects his/her pronunciation of a second language. As accurate and robust the modern mispronunciation detection systems have become, they train HMM models for various mispronunciations using static features, the most popular of which are the Mel-frequency cepstral coefficients (MFCC) which are based on Mel-scale. MFCC are generally thought to provide the best performance \cite{Rabiner} in terms of detection rate. In the text that follows, we introduce a  new ``word adaptive" frequency scale to replace the Mel-scale. From this scale we obtain Adaptive Frequency Cepstral Coefficients (AFCC) and show AFCC have better performance than the conventional MFCC with various commonly used frequency scales.

\section{Feature Optimization for Mispronunciation Detection }

The method of detecting mispronunciation is built on the MFCC-HMM framework, similar in many ways to ASR. In the language learning scenario we address, a student is asked to speak a set of sentences in the foreign language, after which his/her digital recording  is analyzed by the detection software for mispronunciations on a word-by-word basis.  Prior to computing the cepstral coefficients, the short-time frequency scale associated with the speech is warped from a linear to a nonlinear scale, typically the Mel scale. The Mel scale is effective because it results in cepstral features that are better matched to the sensitivity of the human auditory system.  

\begin{figure}  
	\begin{center}  
		\includegraphics[scale = .6]{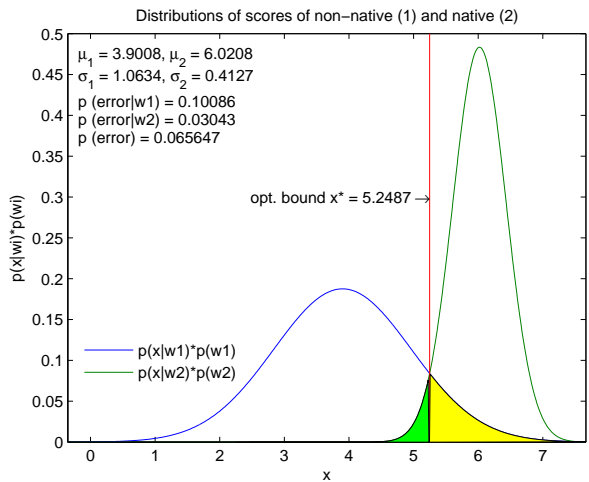}
		\vspace{-0.2 in}
		\caption{\small Distributions of HMM scores of two groups with correct and incorrect pronunciations.
		\label{fig:dist}}  
	\end{center}
	\vspace{-0.2 in}  
\end{figure}

In speech recognition, the Mel-scale warping is generally considered to be the best scale to improve  recognition accuracy. For the application of mispronunciation detection, we introduce adaptivity via a word adaptive frequency scale. A fundamental criterion for comparing the performance between the Mel-scale and the word adaptive frequency scale is the group separation between the native speakers (with correct pronunciation) and the non-native speakers (with incorrect pronunciation) with respect to their distribution of HMM scores. Among the infinite set of frequency scales represented in our construction, the frequency scale with the highest classification rate is considered the optimal scale that provides the best separation for a particular word. The classification rate $r$ is calculated using Bayesian Minimum Error. Assuming the HMM evaluation scores for non-native group ($X_1$) and native group ($X_2$) are Gaussian distributed with 
$X_1 \sim N(\mu_1,\sigma_1^2), X_2 \sim N(\mu_2, \sigma_2^2) $ and $ P(\omega_1) = P(\omega_2) = 1/2 $ where $\omega_{1,2}$ denote the classification of two groups, $P(\mathrm{error}) $ denotes the Bayesian Minimum Error Rate, which can be calculated from the intersection area of the two distributions shown graphically in  Fig. \ref{fig:dist} and calculated using  Eq. (\ref{eql:r}):
\begin{eqnarray} \label{eql:r}
	P(\mathrm{correct}) &=& 1 - (\int_{x^*}^{\infty} p(x \mid \omega_1 ) P(\omega_1 ) \mathrm{d}x  \nonumber \\
                && + \int_{-\infty}^{x^*} p(x\mid \omega_2 ) P(\omega_2 ) \mathrm{d}x )
\end{eqnarray}
where $x^*$ can be found by computing the discriminant function $g(x)$ at $g(x)=0$ where
\begin{equation}
	g(x)=p(x\mid \omega_1 )P(\omega_1 )-p(x\mid \omega_2 )P(\omega_2 )                                      
\end{equation}

\section{Adaptive Frequency Scale}

\subsection{Frequency Scale Generation}

To develop the algorithm, a framework is needed in which the frequency scales can be generated. Recognizing that the scale is bounded between $0$ and the Nyquist frequency, so that we can arguably expect the optimal scale to be one that concentrates its sensitivity in the low frequency region, we first considered a simple $\mu$-law construction.  This approach has the advantage that it can approximate the warping profile we anticipate and it only involves a single parameter. To obtain higher accuracy, however, we have based our algorithm on a dual parameter model (i.e. involving two degrees of freedom) where  Piecewise Cubic Hermite Polynomial (PCHP) interpolation \cite{Fritsch} is used to create the adaptive warping function. 

\begin{figure}[h]
	\begin{center}
		\includegraphics[scale = .6]{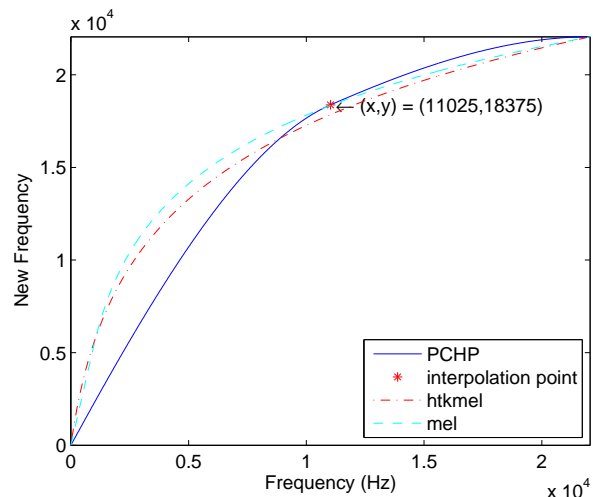} 
 	\end{center}
 	\vspace{-0.2 in}
 	\caption{\small Mel-scales and PCHP interpolated scales. \label{fig:ref-scale}}
\end{figure}

For comparison, we use a) the HTK Mel scale (used in the HTK Speech Recognition Toolkit) \cite{htk} and b) the Mel scale reported by Slaney \cite{Slaney}.\comments{and c) the Bark scale reported by Zwicker \cite{Zwicker}} The Mel and HTK Mel scales essentially employ the same equation and triangular filter banks but differ in their normalization. More specifically, the HTK Mel scale filter banks are normalized to have the same height while the other is normalized to have constant passband area. Fig. \ref{fig:ref-scale} shows HTK Mel and Mel frequency scales compared with an example of the PCHP scale. For convenient display, all frequency plots are normalized to the Nyquist frequency $F_N$ ($F_N=F_s/2=22050$  $\mathrm{Hz}$ in this project, where $F_s$ is the sampling frequency). The PCHP interpolation provides smooth interpolation and a monotonically increasing frequency scale similar to the Mel-scales, but with much more flexibility since the interpolation point used to define the curve can be placed anywhere in the bi-frequency plane. This point is also illustrated in Fig. \ref{fig:ref-scale}.


\subsection{Adaptive Frequency Scale Optimization }
The procedure for PCHP frequency scale optimization can be described as follows: 1) Given any interpolation point $(x_i,y_i)$ in the bi-frequency plane, one can compute the PCHP interpolated frequency scale. 2) Based on that scale, one can then compute the AFCC($x_i,y_i$) for all samples from both native and non-native groups. 3) Using these AFCC($x_i,y_i$) as inputs to the Leave-One-Out HMM training and testing (discussed in Section \ref{system setup}), all samples will be assigned HMM scores $S_{\mathrm{HMM}}(x_i,y_i)$. 4) These HMM scores are then rescaled to the range 1-7 to match the mean and the variance of the human scores  (discussed in Section \ref{database&scoring}). These rescaled scores serve as measurements of pronunciation quality. 5) The classification rate $r(x_i,y_i)$ is computed based on the distribution of the rescaled scores $S_{\mathrm{res}}(x_i,y_i)$ of these two groups using the Bayesian rule.  

Thus, through the process $(x_i,y_i) \rightarrow \mathrm{AFCC}(x_i,y_i) \rightarrow S_{\mathrm{HMM}}(x_i,y_i) \rightarrow S_{\mathrm{res}}(x_i,y_i) \rightarrow r(x_i,y_i)$, any point $(x_i,y_i)$ on the bi-frequency plane can be eventually mapped to a certain classification rate $r(x_i,y_i)$. The frequency scale optimization is an iterative procedure consisting of finding $(x^*,y^*)$,  where

\begin{equation}
  (x^*,y^*) = \underset{x,y\in[0,F_s/2]}{\operatorname{argmax}}(r(x,y))
\end{equation}


\begin{figure}[h]
\begin{minipage}[h]{0.9 \linewidth}
  \centering
  \centerline{\includegraphics[width= 8 cm]{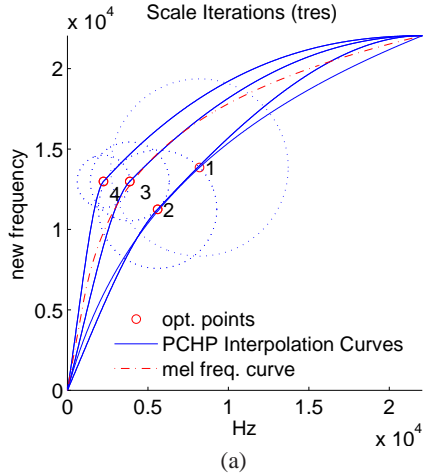}}
  \vspace{-0.05 in}
  \centerline {\small {(a)}}\medskip
\end{minipage}
\hfill
\begin{minipage}[h]{0.9 \linewidth}
  \centering
  \centerline{\includegraphics[width= 8 cm]{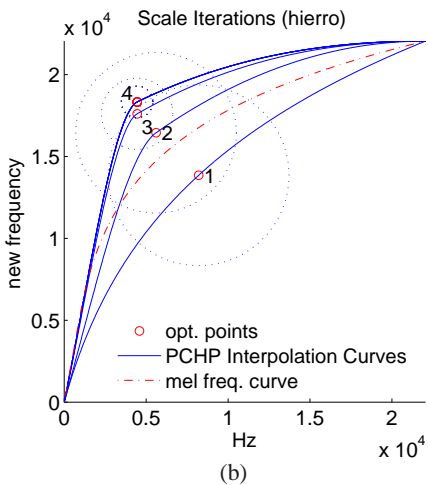}}
  \vspace{-0.05 in}
  \centerline {\small {(b)}}\medskip
\end{minipage}
\vspace{-0.1 in}
\caption{\small Four iterations of frequency scale optimization are shown for the words (a) {\tt tres} and (b) {\tt hierro}.
\label{fig:start-iteration}}  
\end{figure}

\begin{figure}[h]   
\centering
\vspace{-0.1 in}
\includegraphics[scale = .4]{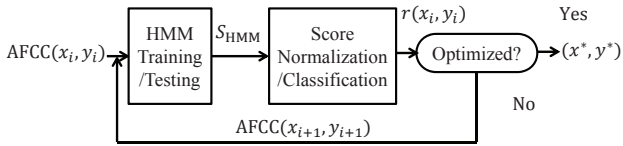} 
\caption{\small Procedure of frequency scale optimization}
\label{fig:procedure}  
\end{figure} 

The above procedure is depicted in Fig. \ref{fig:procedure}. In order to find $(x^*,y^*)$ that maximizes $ r(x,y)$, a modified $N$-step search algorithm which is similar to the 3-step search in motion estimation is employed:

1.	Initialization: a) choose a starting interpolation point $(x_0,y_0)^{(1)}$ in the bi-frequency plane. A good rule of thumb is to use the ``center" of the $\mu$-law scale at $\mu=8$ (the intersection of $\mu$-law scale and the diagonal between the maximum frequencies in both normal and new frequency axes); b) set the search region as a circle centered at $(x_0,y_0)^{(1)}$ with radius $R(1)=\frac{\pi}{2}$ and c) select $M$ candidate searching points $(x_i,y_i)^{(1)}, i = 1,2,...,M$ ($M = 24$ in this project) that are evenly spaced on concentric circles within the search region.

2. Iteration: in each iteration $n$, compute the corresponding classification rate $r(x_i,y_i)^{(n)}$ at each candidate point $(x_i,y_i)^{(n)}$ and set the current optimal classification rate  $r(x^*,y^*)^{(n)}=\max[(r(x_i,y_i)^{(n)}]$.

\begin{itemize}

\item If the current optimal point $(x^*,y^*)^n$ is optimal through all iterations up to $n$, i.e. $(x^*,y^*)^n = {\operatorname{argmax}}(r(x,y))^{(j)},  j = 1,2,...,n)$, then, set a) the next starting point $(x_0,y_0)^{(n+1)}=(x^*,y^*)^{(n)}$ and b) the next search region at the circle centered at $(x_0,y_0)^{(n+1)}$ with radius $R(n+1)=\max[\frac{\pi}{2^{n+1}},|((x_0,y_0)^{(n+1)}- (x_0,y_0)^{(n)}|]$; c) similarly select $M$ candidates $(x_i,y_i)^{(n+1)}, i = 1,2,...,M$, evenly spaced within the new search region;

\item Else, set a counter $k = k+1$ ($k = 0$ initially); keep the starting point, the search region and the search candidates the same in the next iteration, i.e. $(x_0,y_0)^{(n+1)} = (x_0,y_0)^{(n)}$, $R(n+1) = R(n)$ and $(x_i,y_i)^{(n+1)} = (x_i,y_i)^{(n)}$. This repetition compensates for the small variation of HMM scores due to the randomization in HMM training with a relatively small database.
\end{itemize}

3. Termination: if the counter $k = K$, where $K = 3$ in this project, then terminate the iteration. The point providing the largest classification rate is: $(x^*,y^*)= \mathrm{argmax}(r(x^*,y^*)^{(j)},j=1,2,...,n)$.  


Finally, since the radius of the search region $R$ is converging in each  iteration, we are able to find the optimal word adaptive scale (subject to the constraints of the construction) provided by the largest $r(x^*,y^* )$.

\section{Experiments and Results}

\subsection{Database and Human Scoring}
\label{database&scoring}
In this mispronunciation detection project, the talkers are 20 native speakers of English who have completed one yearcollege-level introductory Spanish course and 20 native Spanish students. 10 Spanish words comprise the corpus. Each speaker pronounces each word 10 times. The human scoring juries are composed of 22 adult native speakers of Spanish. Scores range from 1 (poor) to 7 (excellent) based on the level of mispronunciation. In the training of correct and incorrect pronunciation groups, outliers, such as samples from non-native speakers pronounced well enough (closer to the mean score of the native group) or vice versa, are removed so that the samples within each group are more homogeneous.    

\subsection{Optimization System Setup}
\label{system setup}
All samples are noise suppressed and times-scale modified. MFCC/AFCC used in this project are 13 dimensional with 12 cepstral coefficients and 1 energy coefficient. 

The HMMs are trained and tested 5 times using the Leave-One-Out (LOO) algorithm. During each trial, 80\% of the native samples are trained and the remaining 20\% native and all the non-native samples are tested so that every sample has a score (or averaged score).

\subsection{Experimental Results }

\begin{figure}[h]
\vspace{-0.1 in}
\begin{minipage}[h]{\linewidth}
	\begin{center}
		\includegraphics[scale = .6]{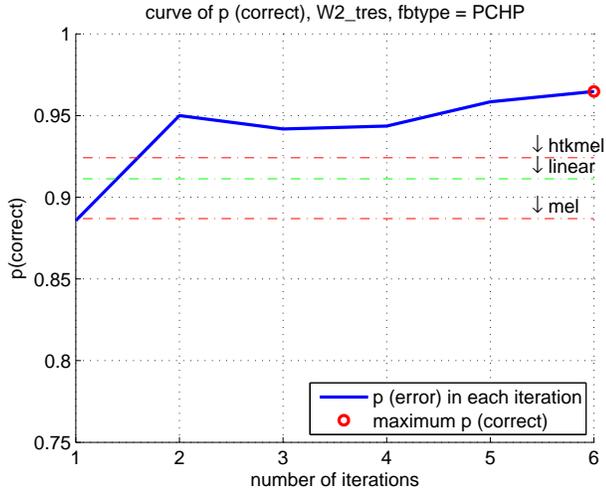} 
 	\end{center}
 	\vspace{-0.15 in}
 	\centerline {\small {(a)}}
 	\vspace{0.05 in}
\end{minipage}
\begin{minipage}[h]{\linewidth}
	\begin{center}
		\includegraphics[scale = .6]{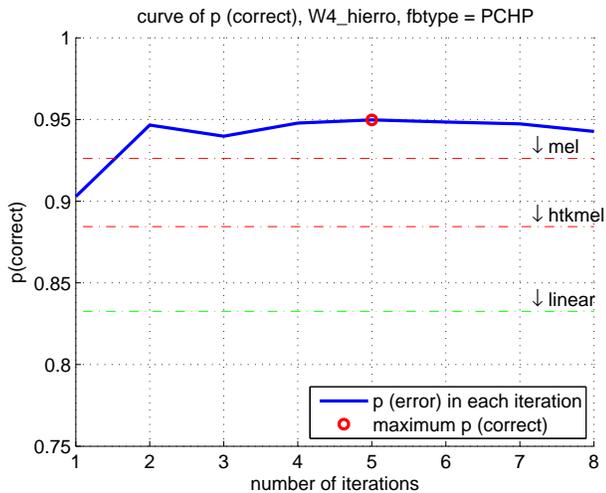} 
 	\end{center}
 	\vspace{-0.15 in}
 	\centerline {\small {(b)}}
 	\vspace{0.05 in}
\end{minipage}
\caption{\small  Classification rates for successive AFCC iterations contrasted against the corresponding rates for the Mel, HTK Mel, Linear scales. The test words used here are {\tt tres} (a) and {\tt hierro} (b).  \label{fig:result1}}
\end{figure}

Fig. \ref{fig:result1} shows the evolution of classification rate $r(x^*, y^*)^{(n)}$ compared with the classification rates associated with Mel, HTK Mel, Bark, and linear scales. The classification rate for the adaptive frequency scale converges very fast and the optimized AFCC outperforms the conventional MFCC systems. The variations that can be seen after the second iteration are due to randomization in the HMM training.

The AFCC result in better frequency scales and better separation of the two groups in terms of classification rate. The improvement is illustrated in Table \ref{table:result3} which shows a comparison of multiple frequency scales performances for 10 Spanish words.


\begin{table}[h]
\begin{small}
\caption{Classification rates for different frequency scales}         
\begin{center}                                                  
\begin{tabular}{l  l  l  l  l  l}                                  
\hline\hline                                        
Scales   & accidente & gemelas & hierro & pala & tres    \\ [0.1 ex]        
\hline                                                        
Adaptive & 0.9571 & 0.9690 & 0.9497 & 0.9374 & 0.9648     \\                         
Mel      & 0.8913 & 0.9159 & 0.9261 & 0.9132 & 0.8869     \\
HTKMel   & 0.9224 & 0.9363 & 0.8844 & 0.9360 & 0.9241     \\
Linear   & 0.9001 & 0.8813 & 0.8326 & 0.8056 & 0.9112     \\
\hline
\end{tabular}
\end{center}
\label{table:result3}
\end{small} 
\vspace{-0.2 in}
\end{table}

\section{Conclusion and Future Work}
In summary, we have introduced an adaptive frequency scale, which yields better separation of native and non-native speakers than the conventional  Mel-scales.  It is possible and likely that even better results can be obtained, since our process was constrained to using a warping function based on a single interpolation point.  Furthermore, if the error surface is irregular, there may be other extreme that result in better solutions.  In future work, we will consider using  multiple points in the PCHP interpolation to increase the degrees of freedom associated with the warping function.  In addition, we will  consider exploiting adaptivity at the syllabic and sub-syllabic levels.

In conclusion, the proposed AFCC have an advantage over conventional MFCC for mispronunciation detection.  The possibility of even greater improvement is possible and a topic of continuing study. 

\bibliography{mypaper_v04}  
  
\end{document}